\documentclass[10pt]{article}
\usepackage{amsmath,amssymb,bm,color,mathtools}
\usepackage[a4paper,left=25mm,top=25mm,]{geometry}

\usepackage[latin9]{inputenc}
\usepackage[american]{babel}
\usepackage{todonotes}
\newcommand{\revise}[1]{\textcolor{black}{#1}}
\usepackage{graphicx}
\usepackage[colorlinks,
            linkcolor=red,
            anchorcolor=blue,
            citecolor=green
            ]{hyperref}

\begin{document}


\title{Discrete Boltzmann model of shallow water equations\\ with polynomial equilibria}
\author{
  Jianping Meng\footnote{Corresponding author.},~Xiao-Jun Gu,~David R Emerson \\
  Scientific Computing Department, STFC Daresbury laboratory, \\Warrington WA4 4AD, United Kingdom\\
  Yong Peng\footnote{Jianping Meng and Yong Peng contributed equally to this work and should be considered as co-first authors.},~Jianmin Zhang \\
State Key Laboratory of Hydraulics and Mountain River Engineering, \\Sichuan University, Chengdu, 610065, P. R. China
}

\maketitle
\begin{abstract}
A type of discrete Boltzmann model for simulating
shallow water flows is derived by using the Hermite expansion approach. Through analytical analysis,  we study the impact of truncating distribution function and discretizing particle velocity space. It is found that the convergence behavior of expansion is nontrivial while the conservation laws are naturally satisfied. Moreover, the balance of source terms and flux terms for steady solutions is not sacrificed. Further numerical validations show that the capability of simulating supercritical flows is enhanced by employing higher order expansion and quadrature.\\
\textbf{Keywords:}
  Discrete Boltzmann model, Shallow water equations, Supercritical flows

\end{abstract}

\section{Introduction}
There is significant interest in modeling shallow water flows
at mesoscopic level \cite{FLD:FLD106,XU2002533,Liang:2007aa,Prestininzi2014439,ZHOU:2002aa,Zhou:2002ab,:ab}, in particular developing the lattice Boltzmann model (LBM) \cite{ZHOU:2002aa,Zhou:2002ab,:ab}. The LBM can be considered as a special discrete velocity method (DVM) for which a minimal discrete velocity set \footnote{Since a discrete velocity set is often obtained from a quadrature for integration,
we consider these two terminologies exchangeable here. In particular,
if the discrete velocities are integer numbers, the terminology
``lattice'' is also used. } is sought and tied to the discretization in space \cite{doi:10.1146/annurev.fluid.30.1.329}. In this way, the LBM is sufficiently simple yet powerful enough to handle complex flow problems, and gradually gains popularity.

A critical challenge for LBM is to simulate supercritical shallow
water flows characterized by a high Froude ($Fr$) number. For this
purpose, an asymmetric model is proposed in \cite{Chopard2013225}
for simulating one-dimensional flows with $Fr>1$. La Rocca et al.~\cite{LaRocca2015117}, has recently proposed a multi-speed model by matching hydrodynamic moments and successfully simulated both one-dimensional and two-dimensional supercritical
flows. Since the discrete velocities are not integer in general,
a finite difference scheme has to be used so that the model is best
classified as a discrete Boltzmann model (DBM) or DVM. However, this model only partially recovers the original simplicity of the LBM.

\revise{The previous works have mainly be achieved by applying the approach of matching hydrodynamic moments~\cite{ZHOU:2002aa,Zhou:2002ab,:ab,LaRocca2015117}. Inspired by these successes, in particular on the supercritical flows~\cite{LaRocca2015117},  we will investigate the Hermite expansion approach \cite{2006JFM550413S,PhysRevLett.80.65} in this work, which is another major approach of seeking optimal DBMs. By using this approach, we are able to derive a range of discrete Boltzmann models with polynomial equilibria (DBMPE). Due to the property of the approach, both integer and non-integer discrete velocity sets can be obtained, and we can have the flexibility of choosing either the stream-collision scheme or general finite difference schemes for simulations. Specifically, the efficient lattice Boltzmann scheme is realised by using the stream-collision scheme. The order of expansion may also be tuned according to the Froude number of the flow. Typically, a second-order expansion is suitable for subcritical flows and a high-order expansion is necessary for supercritical flows. We will discuss the impact of expansion order on the accuracy as well as the well-balancing property of derived DBMs.}

\section{Discrete Boltzmann model with polynomial equilibria}
\subsection{Derivation}
If the vertical length scale is much less than the horizontal length scale, the water
flows are dominated by the nearly horizontal motion and are referred
as shallow water flows. For these flows, the \revise{shallow water equations (SWEs) are} employed
to simplify the modeling, which are written as~\cite{nla.cat-vn2209823}
\begin{equation}
\frac{\partial h}{\partial t}+\bm{\nabla}\cdot(h\bm{V})=0\label{eq:sweh}
\end{equation}
and
\begin{equation}
\frac{D\bm{V}}{Dt}=\frac{\partial \bm{V}}{\partial t}+\bm{V} \cdot \bm{\nabla} \bm{V}=-\frac{1}{h}\bm{\nabla}\mathcal{P}+\bm{F}+\bm{\nabla}\cdot\bm{\sigma},\label{eq:swemo}
\end{equation}
where
\[
\revise{\bm{\nabla} \cdot \bm{\sigma}=\nu \bm{\nabla} \cdot [\bm{\nabla}\bm{V}+(\bm{\nabla}\bm{V})^{T}-(\bm{\nabla}\cdot\bm{V})\bm{\delta}]\thickapprox \nu\triangle\bm{V}},
\]
and $\bm{\delta}$ (i.e., $\delta_{ij}$) is the unit tensor. The equations describe the
evolution of depth, $h$,  and depth-averaged velocity, $\bm{V}=(u,v)$.
The flows are often driven by a body force, $\bm{F}$, which represents not only the effects of an actual body force, including geostrophic
force and tide-raising forces, but also those of wind stress, surface
slope and atmospheric pressure gradient. The pressure
\begin{equation}
\mathcal{P}=\revise{\frac{gh^2}{2}}\label{eq:state}
\end{equation}
originates from the hydrostatic assumption, where $g$ represents for
the gravitational \revise{acceleration}. If necessary, the viscous term $h\nu\triangle\bm{V}$
may also be considered where $\nu$ is the depth averaged \revise{kinematic} viscosity. As has been shown, the SWEs \revise{are} mathematically analogous to the two-dimensional compressible flow equations. In fact, they may be considered
as a system of describing an ideal gas with the state equation Eq. (\ref{eq:state}),
the ratio of specific heat $\gamma=2$, and the ``sound speed'' $\sqrt{gh}$ (cf. Chapter 2.1 in \cite{nla.cat-vn2209823}).

A Boltzmann-BGK type equation can then be constructed for \revise{modeling} SWEs \cite{FLD:FLD106},
\begin{equation}
\frac{\partial f}{\partial t}+\bm{c}\cdot\frac{\partial f}{\partial\bm{r}}+\bm{F}\cdot\frac{\partial f}{\partial\bm{c}}=\frac{f^{eq}-f}{\tau}\label{eq:bkgswe}
\end{equation}
where
\begin{equation}
f^{eq}=\frac{1}{\pi g}\exp[\frac{(\bm{c}-\bm{V})\cdot(\bm{c}-\bm{V})}{gh}].\label{eq:swefeq}
\end{equation}
At the mesoscopic scale, the evolutionary variable becomes the distribution
function $f(\bm{r},\bm{c},t)$ which represents the number of particles
in the volume $d\bm{r}$ centered at position $\bm{r}=(x,y)$ with
velocities within $d\bm{c}$ around velocity $\bm{c}=(c_{x},c_{y})$
at time $t$. The macroscopic quantities can be obtained by integrating
over the whole particle velocity space, i.e.,
\begin{equation}
h=\int fd\bm{c}=\int f^{eq}d\bm{c},\label{eq:h}
\end{equation}
\begin{equation}
h\bm{V}=\int f\bm{c}d\bm{c}=\int f\bm{^{eq}c}d\bm{c,}\label{eq:V}
\end{equation}
and
\begin{equation}
2\mathcal{P}=hgh=\int f(\bm{c}-\bm{V})\cdot(\bm{c}-\bm{V})d\bm{c}.\label{eq:p}
\end{equation}
Moreover, the right-hand side (RHS) term of Eq. (\ref{eq:bkgswe}) obeys the conservation property of the collision integral as shown in Eqs.
(\ref{eq:h}) and (\ref{eq:V}). Similar to an actual ideal gas, the
relation between the relaxation time and kinetic viscosity is
\begin{equation}
\nu h=\mathcal{P}\tau,
\end{equation}
which can be obtained by using the Chapman-Enskog expansion.

In principle, this kinetic equation may be solved directly by using
regular numerical discretization for both physical space and particle velocity space,
i.e., the DVM. Compared with direct
discretization of SWEs, there are two more degrees of freedom for
the particle velocity, which need to be treated carefully for both
accuracy and efficiency. Among various schemes, the \revise{Gauss-type} quadrature
can provide a very efficient yet simple to implement discretization
if properly truncating the equilibrium function Eq. (\ref{eq:swefeq}), i.e., the Hermite expansion approach \cite{CPA:CPA3160020403,2006JFM550413S,PhysRevLett.80.65}. In this work,
we shall derive a type of discrete Boltzmann models for shallow water flows using this approach. For this purpose, it is convenient to first introduce a non-dimensional system,

\begin{equation}
\hat{\bm{r}}=\frac{\bm{r}}{h_{0}},\hat{\bm{V}}=\frac{\sqrt{2}\bm{V}}{\sqrt{gh_{0}}},\hat{t}=\frac{\sqrt{gh_{0}}t}{\sqrt{2}h_{0}},\hat{\bm{F}}=2\frac{\bm{F}}{g}, \label{nonden}
\end{equation}
\[
\hat{\bm{c}}=\frac{\sqrt{2}c}{\sqrt{gh_{0}}},\hat{f}=f\frac{g}{2},\hat{\nu}=\frac{\nu}{\nu_{0}},\hat{h}=\frac{h}{h_{0}},\hat{\mathcal{P}}=\frac{\mathcal{P}}{\mathcal{P}_{0}}=\hat{h}^{2},
\]
where the hat symbol denotes the non-dimensional variables. The reference
depth, $h_{0}$, can be the characteristic
depth of the system such as the initial depth at the inlet (e.g., $h_l$ in Fig.~\ref{fig:ill1D}), while the reference viscosity is represented by $\nu_{0}$ and the reference pressure $\mathcal{P}_0$ is chosen as $gh_{0}^{2}/2$.
By using these non-dimensional variables, Eqs.~(\ref{eq:bkgswe})
and (\ref{eq:swefeq}) become,

\begin{equation}
\frac{\partial\hat{f}}{\partial\hat{t}}+\hat{\mbox{\ensuremath{\bm{c}}}}\cdot\frac{\partial\hat{f}}{\partial\hat{\mbox{\ensuremath{\bm{r}}}}}+\hat{\bm{F\cdot}}\frac{\partial\hat{f}}{\partial\hat{\mbox{\ensuremath{\bm{c}}}}}=\frac{\mathcal{P}_{0}}{\nu_{0}\sqrt{gh_{0}/2}}\frac{\hat{\mathcal{P}}}{\hat{\nu}\hat{h}}(\hat{f}^{eq}-\hat{f}),\label{eq:nondimeq}
\end{equation}
and

\begin{equation}
\hat{f}^{eq}=\frac{1}{2\pi}\exp[\frac{(\hat{\bm{c}}-\hat{\bm{V}})\cdot(\hat{\bm{c}}-\hat{\bm{V}})}{2\hat{h}}],\label{eq:nonswefeq}
\end{equation}
while there is no need to change the form of Eqs.~(\ref{eq:h})~-~(\ref{eq:p}).
We may also define a ``Knudsen'' number

\begin{equation}
\mathcal{K}=\frac{\nu_{0}\sqrt{gh_{0}/2}}{\mathcal{P}_{0}}=\frac{\nu_{0}\sqrt{gh_{0}/2}}{gh_{0}^{2}/2},
\end{equation}
and the local Froude number becomes

\begin{equation}
Fr=\frac{U}{\sqrt{gh}}=\frac{\hat{U}\sqrt{gh_{0}/2}}{\sqrt{g\hat{h}h_{0}}}=\frac{\hat{U}}{\sqrt{2\hat{h}}},\label{eq:Fr}
\end{equation}
where the velocity magnitude is denoted by $U$. Using
the ``Knudsen'' number and the fact that the viscosity
is often considered as a constant, the RHS term of Eq.($\ref{eq:nondimeq}$)
becomes,

\[
\frac{\hat{h}}{\mathcal{K}}(\hat{f}^{eq}-\hat{f}).
\]
Hence, the actual relaxation time may change with time locally. Hereinafter,
we shall use the non-dimensional version of quantities and equations
by default, and the hat symbol will be omitted for clarity.
It is also convenient to use the symbol $\tau$ to substitute
for $\mathcal{K}/\hat{h}$ in writing the equations.

First of all, the equilibrium distribution will be expanded on the basis of the Hermite orthogonal polynomials $\bm{\chi}^{(n)}(\bm{c)}$
in particle velocity space (see Ref. \cite{2006JFM550413S} for detail), i.e.,

\begin{equation}
f^{eq}\approx f_{eq}^{N}=\omega(\bm{c})\sum_{n=0}^{N}\frac{1}{n!}\bm{a}_{eq}^{(n)}\colon\bm{\chi}^{(n)}(\bm{c}),\label{approxfeq}
\end{equation}
where the $N^{th}$ order terms are retained and \revise{the full contraction of tensors are denoted by the symbol $\colon$. The coefficient $\bm{a}_{eq}^{(n)}$ }is given by
\begin{equation}
\bm{a}_{eq}^{(n)}=\int f^{eq}\bm{\chi}^{(n)}d\bm{c} \cong \sum_{\alpha=1}^{d}\frac{w_{\alpha}}{\omega(\bm{c}_{\alpha})}f_{eq}^{N}\bm{\chi}^{n}(\bm{c}_{\alpha}).\label{eq:coeff}
\end{equation}
Using the Gauss-Hermite quadrature with weights $w_{\alpha}$ and abscissae $\bm{c}_{\alpha}$,
$\alpha=1,\cdots,d$, the integration in Eq. (\ref{eq:coeff}) has
been converted into a summation. In particular, the exact equality holds if a sufficient order of quadrature is employed \cite{2006JFM550413S}.  The first few coefficients are
given by
\begin{equation}
a_{eq}^{(0)}=h,\label{eq:a0}
\end{equation}
\begin{equation}
\bm{a}_{eq}^{(1)}=h\bm{V,}
\end{equation}
\begin{equation}
\bm{a}_{eq}^{(2)}=h[\bm{V}^{2}+(h-1)\bm{\delta}],
\end{equation}
\begin{equation}
\bm{a}_{eq}^{(3)}=h[\bm{V}^{3}+(h-1)\bm{\delta}\bm{V}],
\end{equation}
and
\begin{equation}
\bm{a}_{eq}^{(4)}=h[\bm{V}^{4}+(h-1)\bm{\delta}\bm{V}^{2}+(h-1)^{2}\bm{\delta}^{2}],\label{eq:a4}
\end{equation}
where the product of two tensors means the sum of all possible permutations of tensor product (e.g.,  $\bm{\delta}\bm{V}=V_{i}\delta_{jk}+V_{j}\delta_{ik}+V_{k}\delta_{ij}$) while the power of a vector stands for the direct velocity product (e.g., $\bm{V}^3=\bm{V}\bm{V}\bm{V}$).
It is easy to verify that, with an appropriate quadrature, the conservation property of the collision integral will be satisfied automatically
using an expansion higher than the first order, which will simplify
the algorithm. The body force term $\mathcal{F}(\bm{x},\bm{c},t)=-\bm{F}\cdot\nabla_{c}f$
can also be approximated as,
\begin{equation}
\mathcal{F}(\bm{x},\bm{c},t)=\omega(\bm{c})\sum_{n=1}^{N}\frac{1}{(n-1)!}\bm{F}\bm{a}^{(n-1)}\colon\bm{\chi}^{(n)},\label{eq:force}
\end{equation}
where $\bm{a}^{(n)}$ is the corresponding coefficients for the distribution
function, $f$. The first two are same as the $a_{eq}^{0}$ and $\bm{a}_{eq}^{(1)}$
while the higher order terms can be related to stress and heat flux.
Through the expansion, the kinetic equation (\ref{eq:nondimeq}) can
be rewritten in its truncated form, i.e.,
\begin{equation}
\frac{\partial f}{\partial t}+\bm{c} \cdot \frac{\partial f}{\partial \bm{r}}=-\frac{1}{\tau}(f-f_{eq}^{N})+\mathcal{F}.\label{truncatedeq}
\end{equation}
Thus, we will be solving an approximation of the original kinetic
equation.

The second step is to discretize Eq. (\ref{truncatedeq}) in \revise{the} particle
velocity space. The Gauss-Hermite quadrature is \revise{a} natural choice.
In one dimension, the discrete velocities $c_{\alpha}$ are just the
roots of Hermite polynomials, and the corresponding weights are
determined by:
\begin{equation}
w_{\alpha}=\frac{n!}{[n\chi^{n-1}(c_{\alpha})]^{2}}.\label{weight}
\end{equation}
Given one-dimensional velocity sets, those of a higher-dimension
can be constructed using the \lq\lq production\rq\rq~formulae
\cite{2006JFM550413S}. Once the discrete velocity set is chosen,
the governing equation is discretized as
\begin{equation}
\frac{\partial f_{\alpha}}{\partial t}+\bm{c}_{\alpha}\cdot\frac{\partial f_{\alpha}}{\partial\bm{r}}=-\frac{1}{\tau}\left(f_{\alpha}-f_{\alpha}^{eq}\right)+\mathcal{F_{\alpha}},\label{lbgk}
\end{equation}
where $f_{\alpha}=w_{\alpha}f(\bm{r},\bm{c}_{\alpha},t)/\omega(\bm{c}_{\alpha})$,
$f^{eq}_\alpha=w_{\alpha}f^{N}_{eq} (\bm{r},\bm{c}_{\alpha},t)/\omega(\bm{c}_{\alpha})$
and $\mathcal{F}_{\alpha}=w_{\alpha}\mathcal{F}(\bm{r},\bm{c}_{\alpha},t)/\omega(\bm{c}_{\alpha})$.
According to Eqs. (\ref{eq:a0}) - (\ref{eq:a4}) and (\ref{approxfeq}),
the explicit form of the fourth order, $f_{\alpha}^{eq}$, is
\begin{equation}
\begin{aligned}
f_{\alpha}^{eq}&= hw_{\alpha}(1 + c_{i}V_{i}+\frac{1}{2}((c_{i}V_{i})^{2}-V_{i}V_{i}+(h-1)(c_{i}c_{i}-2))\\
&\underbrace{
 +\frac{c_{i}V_{i}}{6}((c_{i}V_{i})^{2}-3V_{i}V_{i}+3(h-1)(c_{i}c_{i}-4))}_\text{3rd} \\
& \begin{rcases*}
+\frac{1}{24}((c_{i}V_{i})^{4}-6(V_{i}c_{i})^{2}V_{j}V_{j}+3(V_{j}V_{j})^{2})\\
+\frac{h-1}{4}((c_{i}c_{i}-4)((V_{i}c_{i})^{2}-V_{i}V_{i})-2(V_{i}c_{i})^{2}) \\
+ \frac{(h-1)^{2}}{8}((c_{i}c_{i})^{2}-8c_{i}c_{i}+8))
 \end{rcases*}\text{4th} \\
 \end{aligned},
 \end{equation}
\revise{where the third order and the fourth order terms are highlighted.}

We have now obtained the DBMPE (\ref{lbgk}). To conduct a numerical simulation, however, the physical space $\bm{r}$ and
the time $t$ are needed to be discretized. For this purpose, any available
schemes may be utilized
according to the property of flow, such as finite difference/volume/element schemes. In particular, if using abscissae consisting
of integers, the lattice Boltzmann scheme
\begin{equation}
 \tilde{f}_{\alpha}(\bm{r}+\bm{c}_{\alpha}dt,t+dt)-\tilde{f}_{\alpha}(\bm{r},t)=
  -\frac{dt}{\tau+0.5dt}\left[\tilde{f}_{\alpha}(\bm{r},t)-f_{\alpha}^{eq}(\bm{\bm{r}},t)\right]+\frac{\tau\mathcal{F}_{\alpha}dt}{\tau+0.5dt}\label{eq:scheme}
\end{equation}
can be constructed by introducing

\[
\tilde{f}_{\alpha}=f_{\alpha}+\frac{dt}{2\tau}(f_{\alpha}-f_{\alpha}^{eq})-\frac{dt}{2}\mathcal{F}_{\alpha},
\]
which allows the stream-collision scheme. Due to this
unique property, the LBM has attracted significant interests in broad areas including shallow water simulations. However,
if a high-order expansion is used, the abscissae will not be integer
in general so that more sophisticated schemes will be required to
solve the DBMPE.

Now we are \revise{ready} give some remarks on the connections of the DBMPE (\ref{lbgk}), the LBM (\ref{eq:scheme}) and a general DVM. Essentially, the DBMPE is a specific DVM with carefully chosen discretization in particle velocity space and expansion order for distribution functions so that the conservation laws are satisfied precisely, and then the LBM is a DBMPE with a fixed numerical discretizations Eq.~(\ref{eq:scheme}) in the physical space $\bm{r}$ and time $t$. In general, the numerical discretizations in $\bm{r}$ and $t$ are not specified for the DBMPE, any available discretization \revise{techniques in space and time} can be applied for Eq.~(\ref{lbgk}) including finite difference/volume/element schemes, which are correspondingly named as finite difference/volume/element LBM in community (see also \cite{LaRocca2015117}). In fact, Eq.~(\ref{eq:scheme}) is a special finite difference scheme. Nevertheless, the DBMPE maintains an important feature of the LBM, i.e., seeking minimal discretization set in the \revise{particle} velocity space which is guided by satisfying conservation property\cite{doi:10.1146/annurev.fluid.30.1.329}. Compared with a general DVM, the computational cost \revise{is} often be reduced.  In this work, we focus on the derivation of DBMPE (\ref{lbgk}) and the impact of truncation of Hermite expansion and the relevant Gauss-Hermite quadrature on the capability for simulating shallow water flows, in particular, supercritical flows.

\subsection{Remarks on accuracy of equation (\ref{lbgk})}
\label{secaccuracy}
As previously shown, errors are introduced to the DBMPE (\ref{lbgk}) by truncating equilibrium
function and discretizing particle
velocity space, i.e., from Eq. (\ref{eq:nondimeq}) to Eq. (\ref{truncatedeq})
and from Eq. (\ref{truncatedeq}) to Eq. (\ref{lbgk}).

In principle, if the order of the Hermite expansion is sufficiently high,
Eq. (\ref{truncatedeq}) is expected to accurately recover Eq. (\ref{eq:nondimeq}).
In practice, however, only a few orders may be affordable in term of computational cost, and the
approximation accuracy will be determined by the expansion order.
In general, the accuracy will be related to the Froude number,
i.e., higher Froude numbers may need more expansion terms, cf. Eq.~(\ref{eq:nonswefeq}) and Eq.~(\ref{eq:Fr}). For instance,
similar to gas dynamics, see Ref. \cite{Meng2011b}, using a first-order expansion leads to a linear equation which is \revise{only} suitable
when $Fr\rightarrow0$. Specifically, we may introduce a $L^{2}$
norm
\[
E_{T}=\text{\ensuremath{\sqrt{\frac{\int(f^{eq}-f_{eq}^{N})^{2}d\bm{c}}{\int(f^{eq})^{2}d\bm{c}}}}}
\]
\begin{figure*}[hbtp!]
\begin{center}
\includegraphics[width=0.45\textwidth]{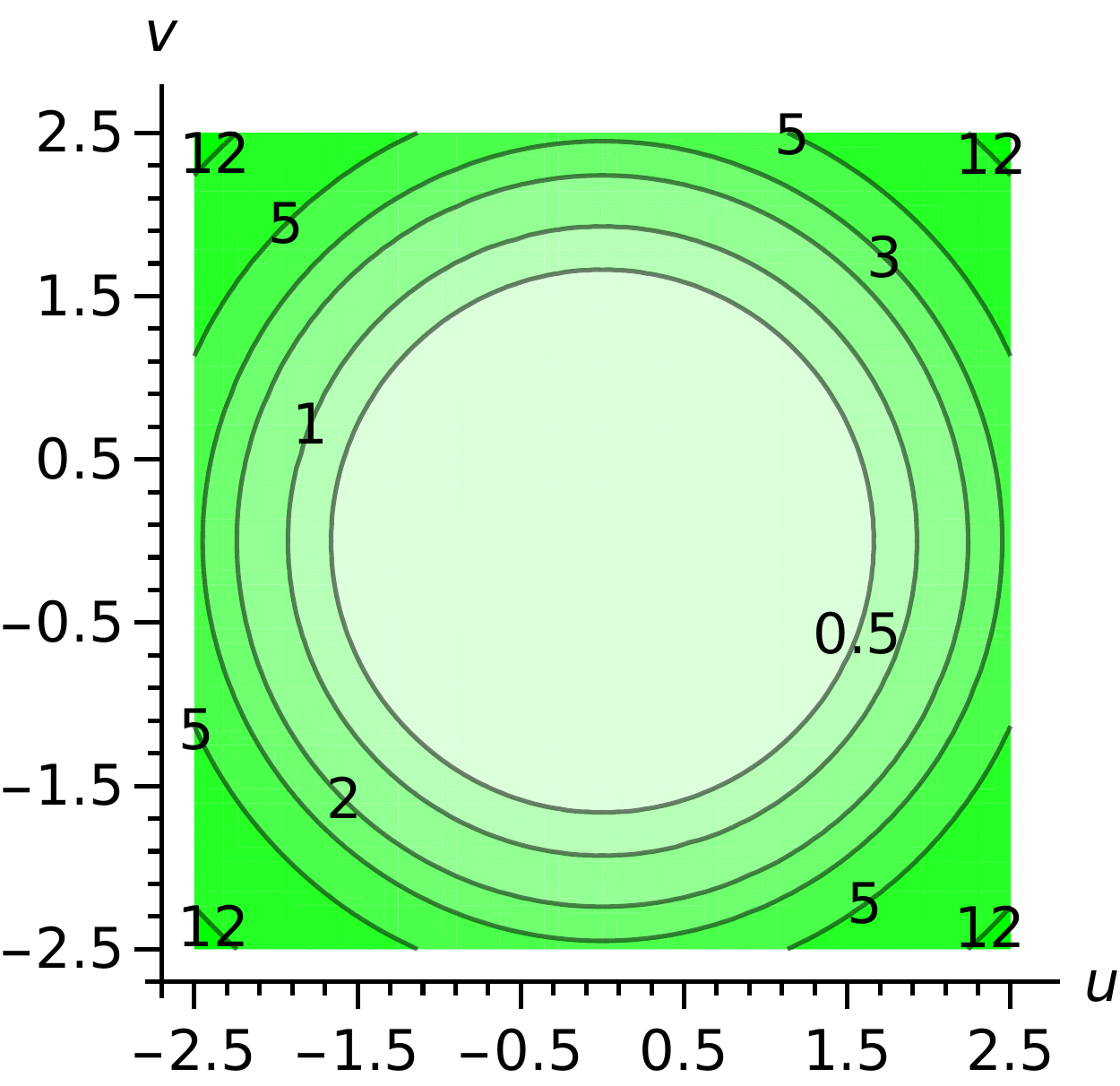}\hspace{1cm}
\includegraphics[width=0.45\textwidth]{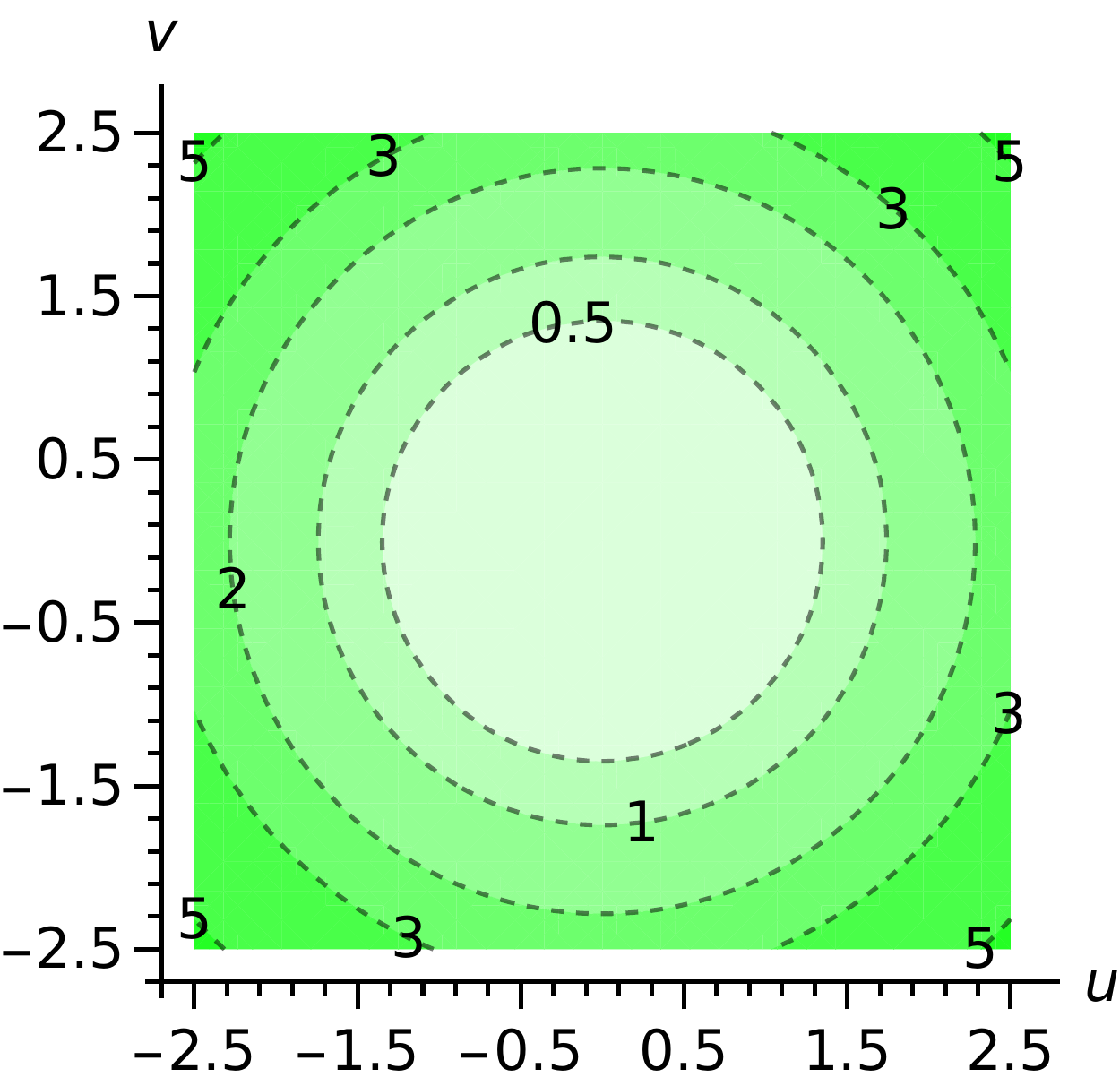}\\
\includegraphics[width=0.45\textwidth]{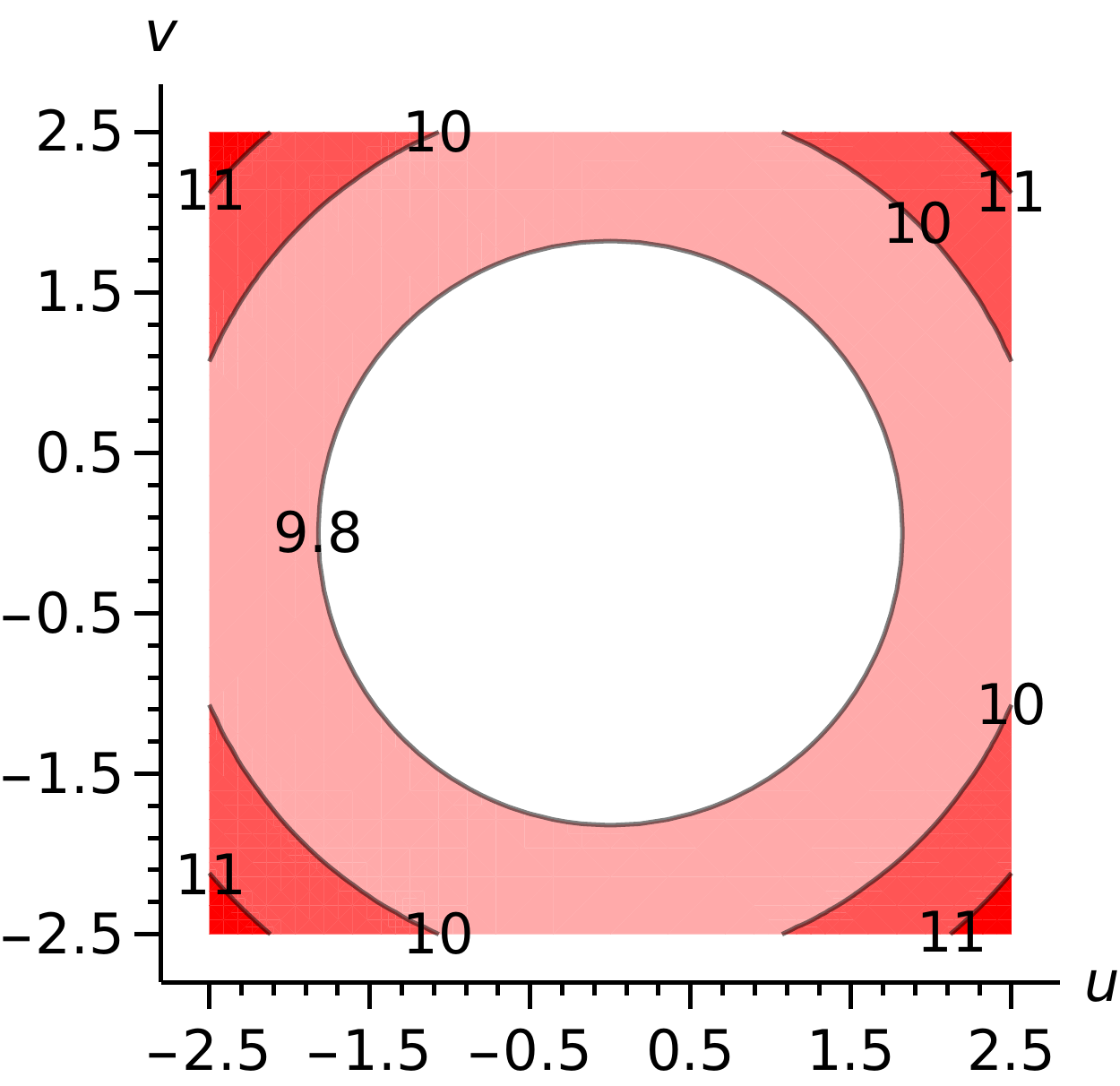}\hspace{1cm}
\includegraphics[width=0.45\textwidth]{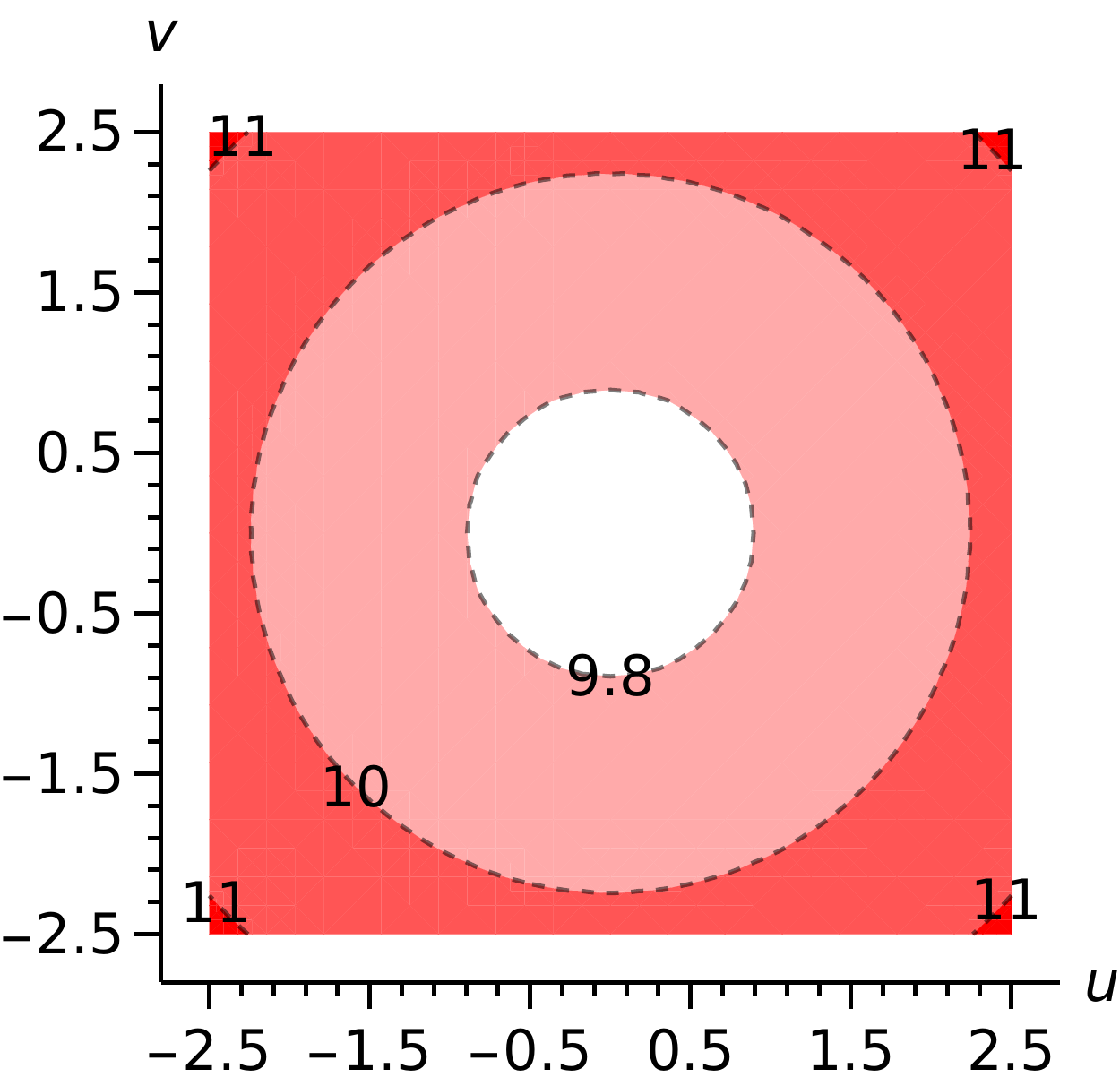}
\caption{\revise{Contours of $E_T$ when the fourth-order (left, solid contour lines) and  the second-order (right, dashed contour lines) expansion are used respectively for $f_{eq}^{N}$ at $h=1$ (top, green color gradient) and $h=0.001$ (bottom, red color gradient). To better display the contour labels at $h=0.001$, the $E_T$ values are transformed by $100(E_T-0.9)$, which leads to the relationships, $9.8 \rightarrow E_T=0.998$, $10 \rightarrow E_T=1$ and $11 \rightarrow E_T=1.01$}.  \label{fig:ET}} \end{center}
\end{figure*}to measure the \revise{difference} of $f_{eq}^{N}$ from $f^{eq}$ over the
whole particle velocity space. In Fig.~\ref{fig:ET}, \revise{contours of $E_{T}$ over the macroscopic velocity range of
$u\in\{-2.5,2.5\}$ and $v\in\{-2.5,2.5\}$ are plotted}. \revise{For comparison, we consider both the second-order expansion and fourth-order expansion of the equilibrium
distribution function. It is also worthwhile to note that $E_T$ represents an accumulation of errors from all discrete velocities so that its value can be large. From these contours, we clearly see that the error
grows with increasing macroscopic velocities in general. In comparsion to the second-order expansion, the fourth-order expansion leads to smaller error where the speed is relatively small but increases the error quickly with larger speed. When $h=0.001$, the error grows slower but its magnitude is farily significant (roughly $1$ for both orders of expansions) even with the zero speed. Overall, the results imply that it is challenge for a DBM to simulate flows with large speed and large depth ratio.} However,
since the conservation laws are retained naturally, even with limited expansion
order, the DBMPE may be sufficient for a broad range of the hydrodynamic
problems, particularly if the diffusion term plays a negligible role~\revise{(e.g., the numerical tests in Sec.~\ref{numervali})}.

The second kind of error is from the chosen quadrature. According
to the Chapman-Enskog expansion, when the ``Knudsen'' number $\mathcal{K}$
is small, the first-order asymptotic solution of Eq. (\ref{truncatedeq})
may be written as

\begin{equation}
f^{(1)}=-\tau(\partial_{t}^{(0)}+\bm{c}\cdot\bm{\nabla}+\bm{F}\cdot\bm{\nabla}_{c})f^{(0)},\label{eq:1stsol}
\end{equation}
where the zeroth-order solution, $f^{(0)}$, is just the truncated equilibrium
function, $f_{eq}^{N}$. From Eq.~(\ref{eq:1stsol}), $f^{(1)}$ will
include a polynomial of $\bm{c}$ of one order higher than that in
$f^{(0)}$. Therefore, to numerically evaluate the integration of
Eqs. (\ref{eq:h}) to (\ref{eq:p}), we need a quadrature with the
degree of precision
\[
P=2d-1\geq N+1+M,
\]
to calculate the $M^{th}$ order of moment (e.g., $M=0$ for the zeroth-order moment, $h$). Specifically, if using a $4^{th}$ order Hermite
expansion and to get the stress right ($M=2$), we need at least a
quadrature with the $7^{th}$ order of precision, which may need four
discrete velocities in one dimension.
\subsection{Remarks on external force term}
For modeling shallow water flows, a  well-balanced numerical scheme is often desired, which precisely satisfies the steady SWEs at discrete level \cite{NoelleHighorderWellbalancedSchemes2010,XU2002533}. For the brevity of discussion, the Euler type equations are often employed in literature, i.e.,
\begin{equation}
\bm{\nabla} \cdot (h \bm{V})=0,
\end{equation} and
\begin{equation}
\bm{\nabla} \cdot (h\bm{V}^2)+ \bm{\nabla} \mathcal{P}= h \bm{F}\label{eq:eulerm},
\end{equation} where the variables are already normalized using Eq.~(\ref{nonden}).
In this sense, a well-balanced scheme means that the discretization for the RHS and LHS terms of Eq.~(\ref{eq:eulerm}) will be carefully chosen to ensure the validity of Eq.~(\ref{nonden}) in its discrete form

Here we mainly concern if the truncation on the Hermite expansion and the discretization in \revise{the} particle velocity space can satisfy the well-balanced property. In other words, if Eq.~(\ref{lbgk}) can recover Eq.~(\ref{eq:eulerm}) under the steady state. For this purpose, we substitute Eq.~(\ref{approxfeq}) and Eq.~(\ref{eq:force}) into Eq.~(\ref{lbgk}), multiply both RHS and LHS terms by $\bm{c}$, and integrate them over the particle velocity space. Thus we need to validate if the equation
\begin{equation}
  \bm{\nabla} \cdot \int f_{eq}^N \bm{c}^2 d\bm{c}=\int \mathcal{F} \bm{c} d\bm{c},
\end{equation} is equivalent to Eq.~(\ref{eq:eulerm}). We note that the integration and the summation are exchangeable in such as Eq.~(\ref{eq:coeff}) with sufficiently accurate quadrature. Considering Eqs.~(\ref{eq:coeff}), (\ref{eq:a0})-(\ref{eq:a4}) and  (\ref{eq:force}), it is straightforward to obtain that
\begin{equation}
  \int f_{eq}^N \bm{c}^2 d\bm{c}= h\bm{V}^2+\mathcal{P} \bm{\delta},
\end{equation} and
\begin{equation}
  \int \mathcal{F} \bm{c} d\bm{c}= h \bm{F}.
\end{equation} Therefore, the discretization in particle velocity space precisely holds the well-balanced property.

For a practical simulation, Eq.~(\ref{lbgk}) has to be further discretized in $\bm{r}$ and $t$ and  becomes algebraic equations e.g., Eq.~(\ref{eq:scheme}), which may alter the well-balanced property. As we have no preference for such discretizaions in this work, we will leave the discussion open for future works where the relevant schemes are specified. In fact, a \revise{few} different schemes will be employed in the following to serve the purpose of studying the impact of the Hermite expansion and the corresponding Gauss-Hermite quadrature. Nevertheless, it is possible to realize the well-balanced property for the standard stream-collision scheme, as shown \revise{by Zhou} \cite{:ab}.
\section{Numerical validation}
\label{numervali}
In the following, we will numerically validate if the capability of simulating supercritical flows can be enhanced by using higher order expansion and quadrature. For this purpose, we first need to determine the discretization scheme in space $\bm{r}$ and $t$. Since their impacts are not our focus in this work, we employ \revise{a few simple schemes according the type of Hermite abscissae, i.e., the combination of upwind schemes and an Euler forward scheme in time for general discrete velocity sets, and the scheme Eq.~(\ref{eq:scheme}) for integer velocities. Afterwards, the fully discretized equations are solved for two classical problems, i.e., one-dimensional dam-\revise{break} problem and two-dimensional circular dam-\revise{break} problem.}

By utilizing Eq.~(\ref{approxfeq}), the
Hemite abscissae and Eq.~(\ref{weight}), it is straightforward to
construct various discrete velocity sets. For instance, to simulate high Froude number flows, we may need to use a fourth
order expansion for $f_{\alpha}^{eq}$ and a quadrature of at least
$7^{th}$ order of precision, see discussions in Sec.~\ref{secaccuracy}. By contrast, if the flow is subcritical,
then it may be enough to employ commonly used \revise{nine-velocity set} and a second order expansion for $f_{\alpha}^{eq}$. When the fourth
expansion terms are employed for $f_{\alpha}^{eq}$, we use a set of 16 discrete velocities obtained
from the roots of fourth order Hermite polynomial. These 16 discrete
velocities are not integer and require a finite difference scheme. For convenience, \revise{we list the roots of the fourth-order Hermite polynomial in one-dimensional space in} Table \ref{tab:d2q16}. The two-dimensional 16-velocity set can then be constructed by using the  \lq\lq production\rq\rq formulae.  The two dimensional 9-velocity set is commonly used so that the detail is ignored here.
\begin{table}[ht]
\caption{\revise{Quadrature from the fourth-order Hermite polynomial. The two-dimensional set can be constructed by using the  \lq\lq production\rq\rq formulae}, i.e., $\bm{c}_\beta$=($c_i$,$c_j$) and $w_\beta=w_i w_j$ where $\beta \in \{0..15\}$ and $i$,$j \in \alpha$.}
\begin{center}
{\begin{tabular}{|p{2cm}||p{5cm}|p{5cm}|}\hline
$\alpha$ & Roots ($c_\alpha$) & Weights ($w_\alpha$)\\
 \hline
 0   & $-\sqrt{3-\sqrt{6}}$    & $\frac{12}{\left(6 \sqrt{2\left(3-\sqrt{6}\right)}-2\sqrt{2}\left(3-\sqrt{6}\right)^{3/2}\right)^2}$\\
 1 &  $\sqrt{3-\sqrt{6}}$  & $\frac{12}{\left(6 \sqrt{2\left(3-\sqrt{6}\right)}-2\sqrt{2}\left(3-\sqrt{6}\right)^{3/2}\right)^2}$  \\
 2 &$-\sqrt{3+\sqrt{6}}$ & $\frac{12}{\left(6 \sqrt{2\left(3+\sqrt{6}\right)}-2\sqrt{2}\left(3+\sqrt{6}\right)^{3/2}\right)^2}$\\
 3    &$\sqrt{3+\sqrt{6}}$ & $\frac{12}{\left(6 \sqrt{2\left(3+\sqrt{6}\right)}-2\sqrt{2}\left(3+\sqrt{6}\right)^{3/2}\right)^2}$\\
 \hline
\end{tabular} \label{tab:d2q16}}
\end{center}
\end{table}

\begin{figure}[hbtp!]
\begin{center}
\includegraphics[width=0.9\textwidth]{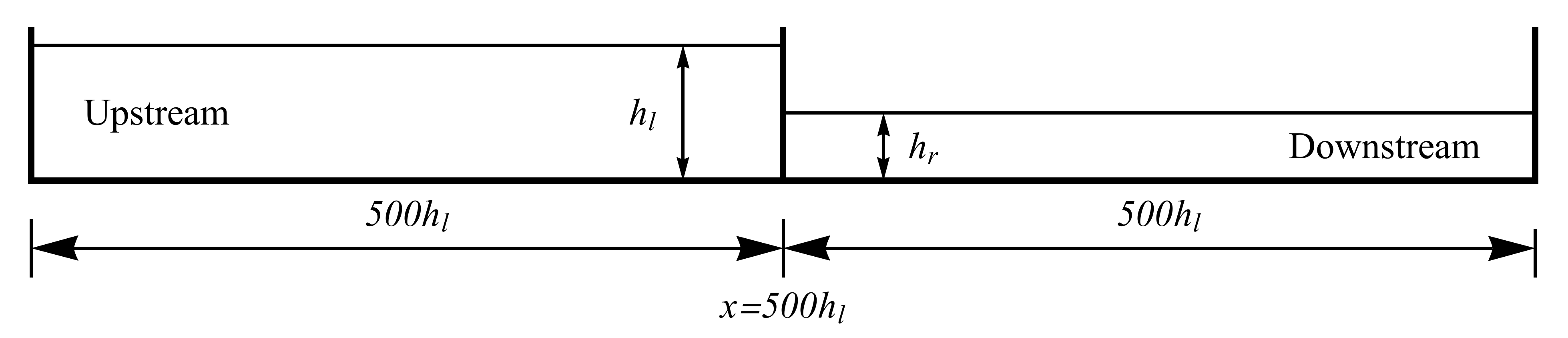}
\caption{\revise{Configuration of the one-dimensional dam-break problem at the initial time (not to scale)}.\label{fig:ill1D}}
\end{center}
\end{figure}

For the one-dimensional dam-\revise{break} problem, we set the whole domain to be $1000$ in the
non-dimensional system (i.e., $1000 h_l$ in the dimensional system while the initial
depth $h_l$ at the inlet is chosen as the reference length). An initial discontinuity will be placed at the
middle $x=500$ (see Fig.~\ref{fig:ill1D}). For supercritical flow, the initial water \revise{depth
is $1$ at the left half domain (i.e., the reference length scale) and $0.001$ at the right
half}, while for the subcritical case, the ratio is $1:0.25$.
\begin{figure*}[hbtp!]
\begin{center}
\includegraphics[width=0.4\textwidth]{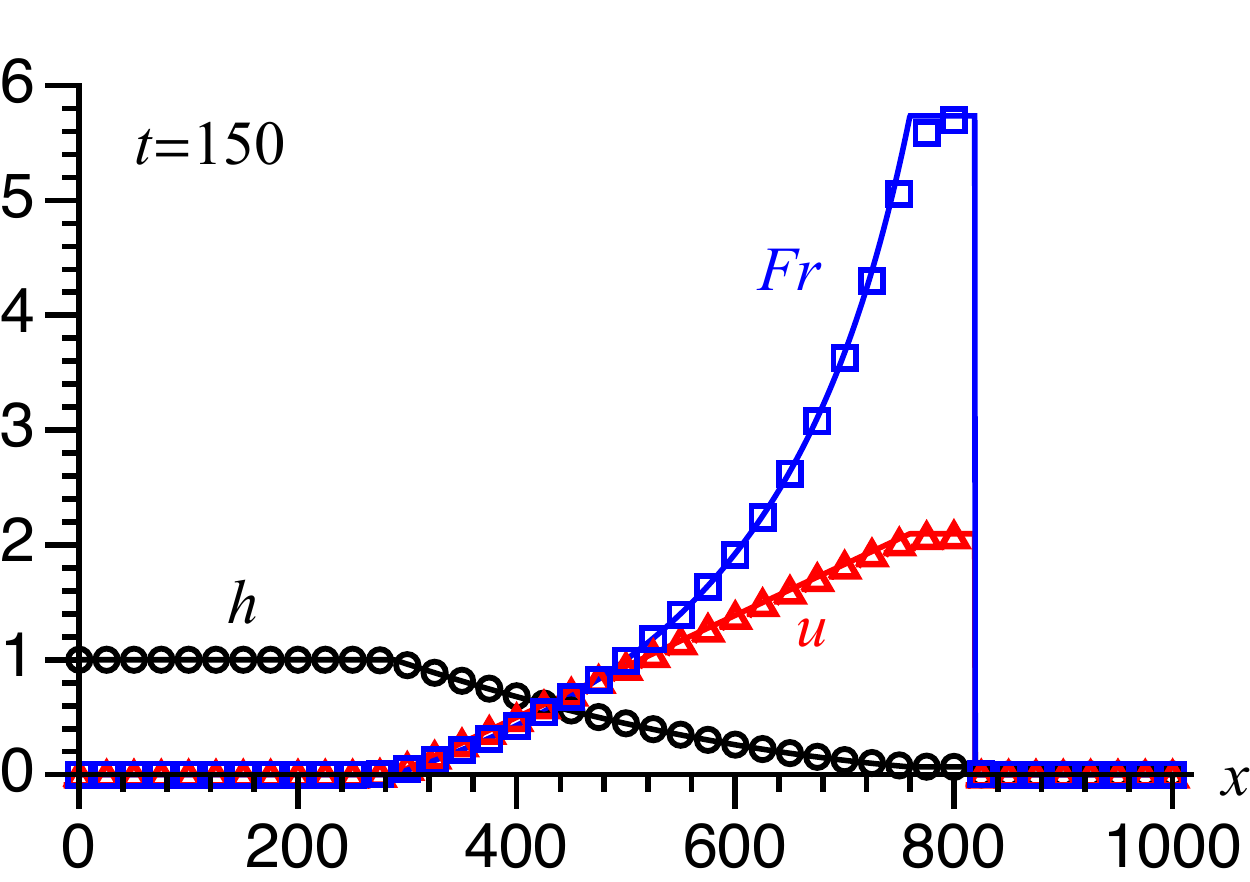}\hspace{1cm}
\includegraphics[width=0.4\textwidth]{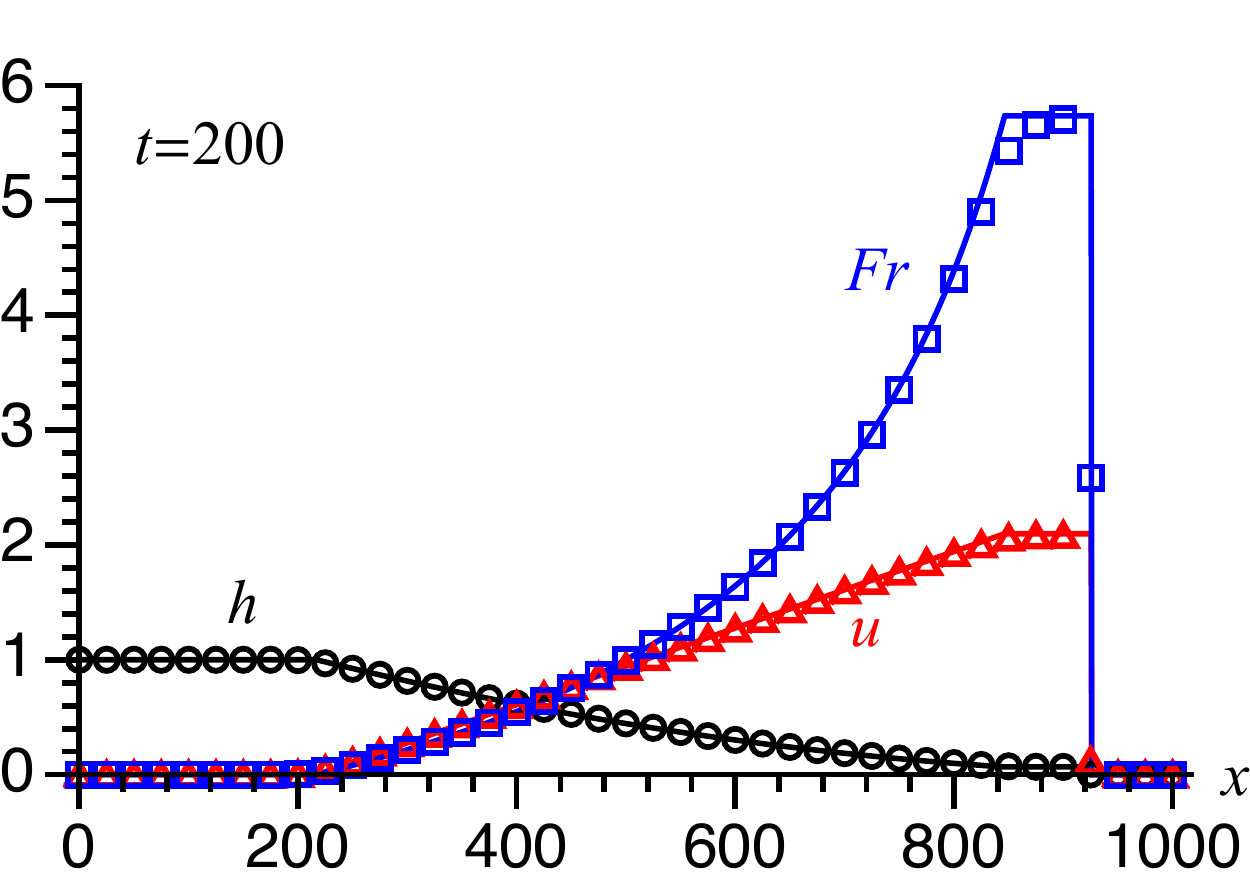}
\caption{Profiles of depth, velocity and local Froude number for the one-dimensional dam-\revise{break} problem with  $h_l:h_r=1000:1$. Lines and symbols represent analytical and numerical solutions, respectively. \label{fig:1000vs1}}
\end{center}
\end{figure*}
\begin{figure*}[hbtp!]
\begin{center}
\includegraphics[width=0.4\textwidth]{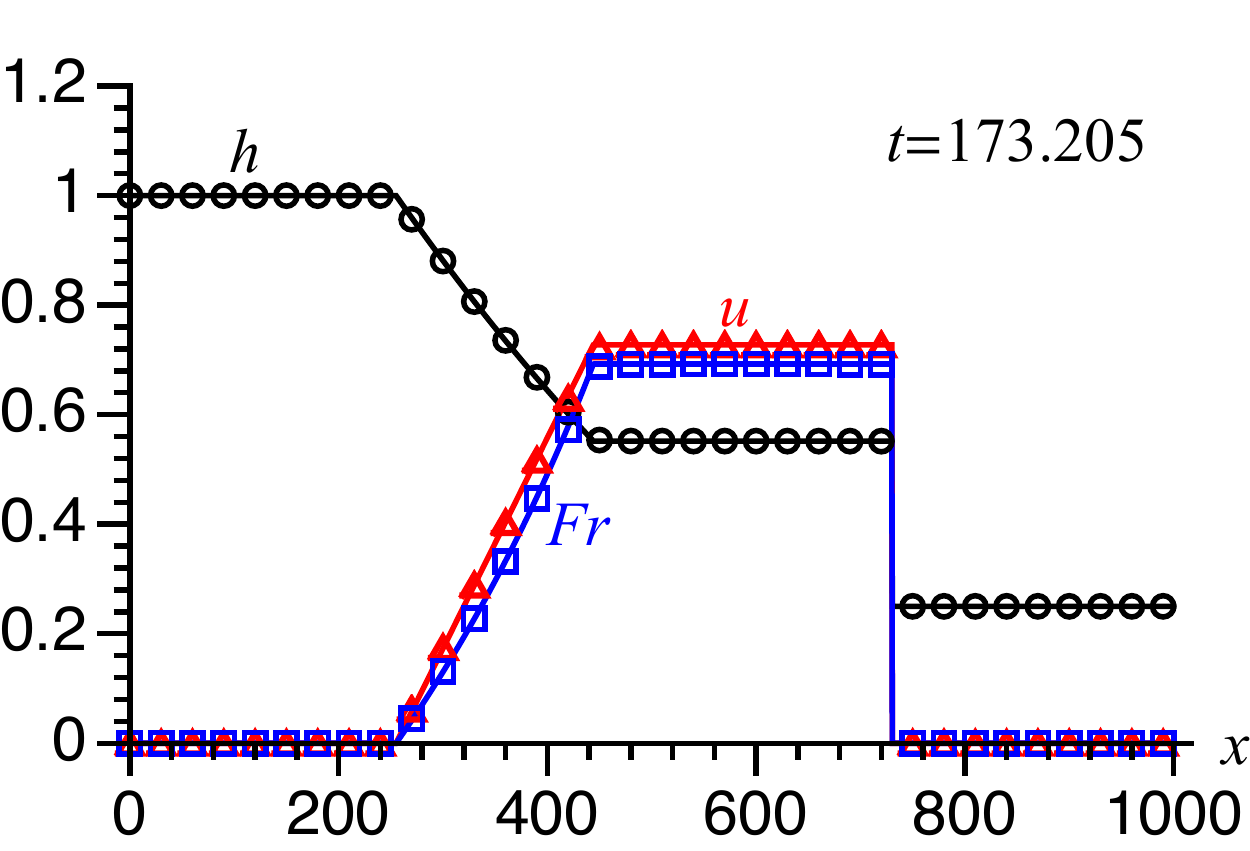}\hspace{1cm}
\includegraphics[width=0.4\textwidth]{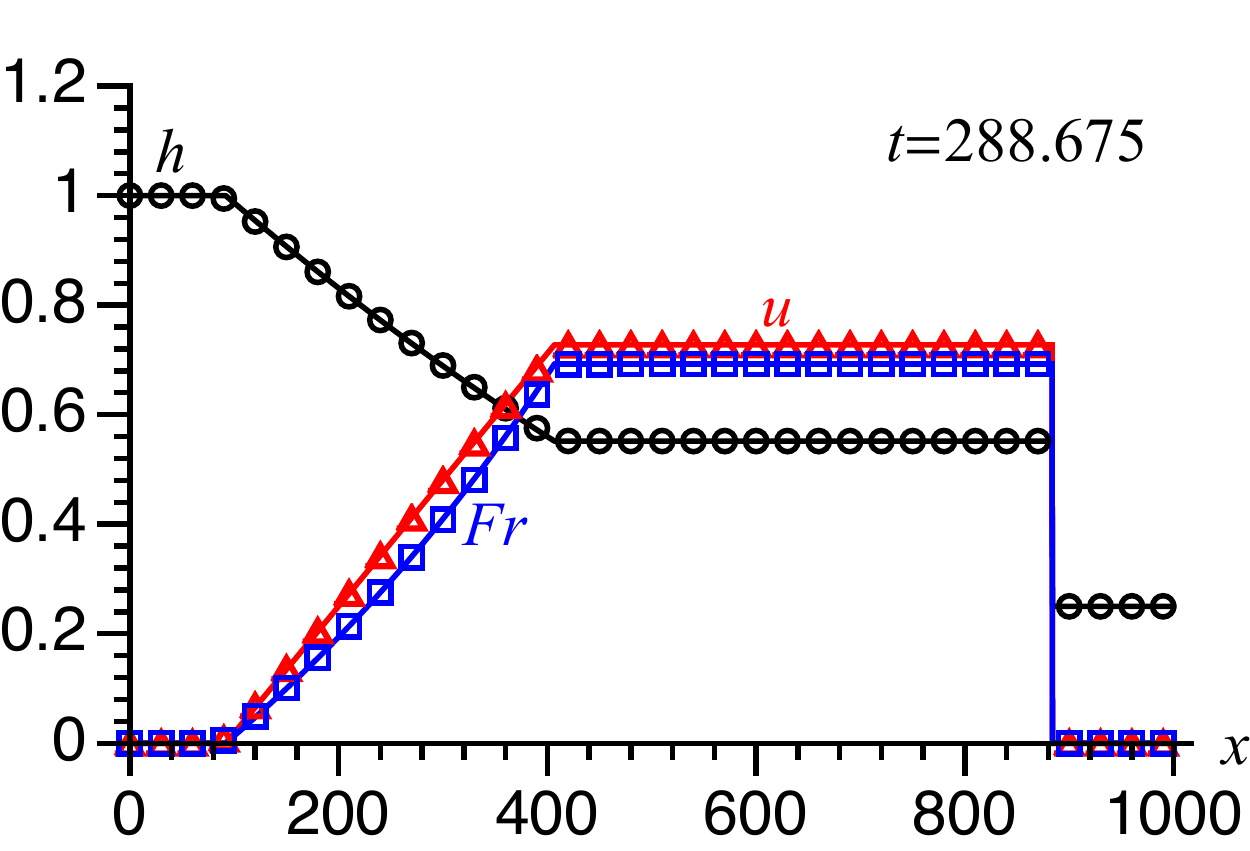}
\caption{Profiles of depth, velocity and local Froude number for the one-dimensional dam-\revise{break} problem with $h_l:h_r=4:1$. Lines and symbols represent analytical and numerical solutions, respectively.\label{fig:4vs1}}
\end{center}
\end{figure*}

\revise{To test the supercritical flow}, the parameters for calculation are chosen
as $\mathcal{K}=0.001$, $dx=0.5$, and $dt=0.0001$, and the 16-velocity set is employed with the fourth-order expansion. \revise{Since the quadrature is non-integer, a first-order upwind schems is chosen for the space discrtization.} The results at $t=150$ and $t=200$ are shown in Fig.~\ref{fig:1000vs1}.
Excellent agreement is found for both water depth and velocity with
the analytical solution~\cite{delestre:hal-00628246}, although numerical error is found for the profile of local Froude number at the position of discontinuity
due to the phase lag between water depth and velocity caused by numerical
diffusion of upwind schemes~\cite{LaRocca2015117}. The simulations confirms the capability of modeling supercritical
flows with a fourth order expansion and corresponding quadrature.
By contrast, we also tested the combination of a second-order expansion and 16 discrete velocities, but found that the
simulation is unstable using the same problem setup.

For the subcritical flow, we use the stream-collision scheme (\ref{eq:scheme})
and the parameters are chosen as
$\mathcal{K}=0.1$, $dx=0.5$, and $dt=dx/\sqrt{3}$. The numerical results are shown in Fig.~\ref{fig:4vs1}, where excellent
agreement with the analytical solutions~\cite{delestre:hal-00628246} can be found. Moreover, it appears that there is almost no numerical diffusion, which is consistent with the property of the scheme (\ref{eq:scheme}). However, possibly due to this property,
the capability in simulating supercritical flows is limited. By using a multispeed lattice with 37 integer discrete velocities together with a fourth order expansion for the equilibrium function, we observe instability when the ratio is  $h_l:h_r=20:1$ using the stream-collision scheme, which confirms the findings of Peng et al.\cite{Peng:2017aa} However, an appropriate finite difference scheme can still maintain stable simulations.

\begin{figure}[hbtp!]
\begin{center}
\includegraphics[width=0.45\textwidth]{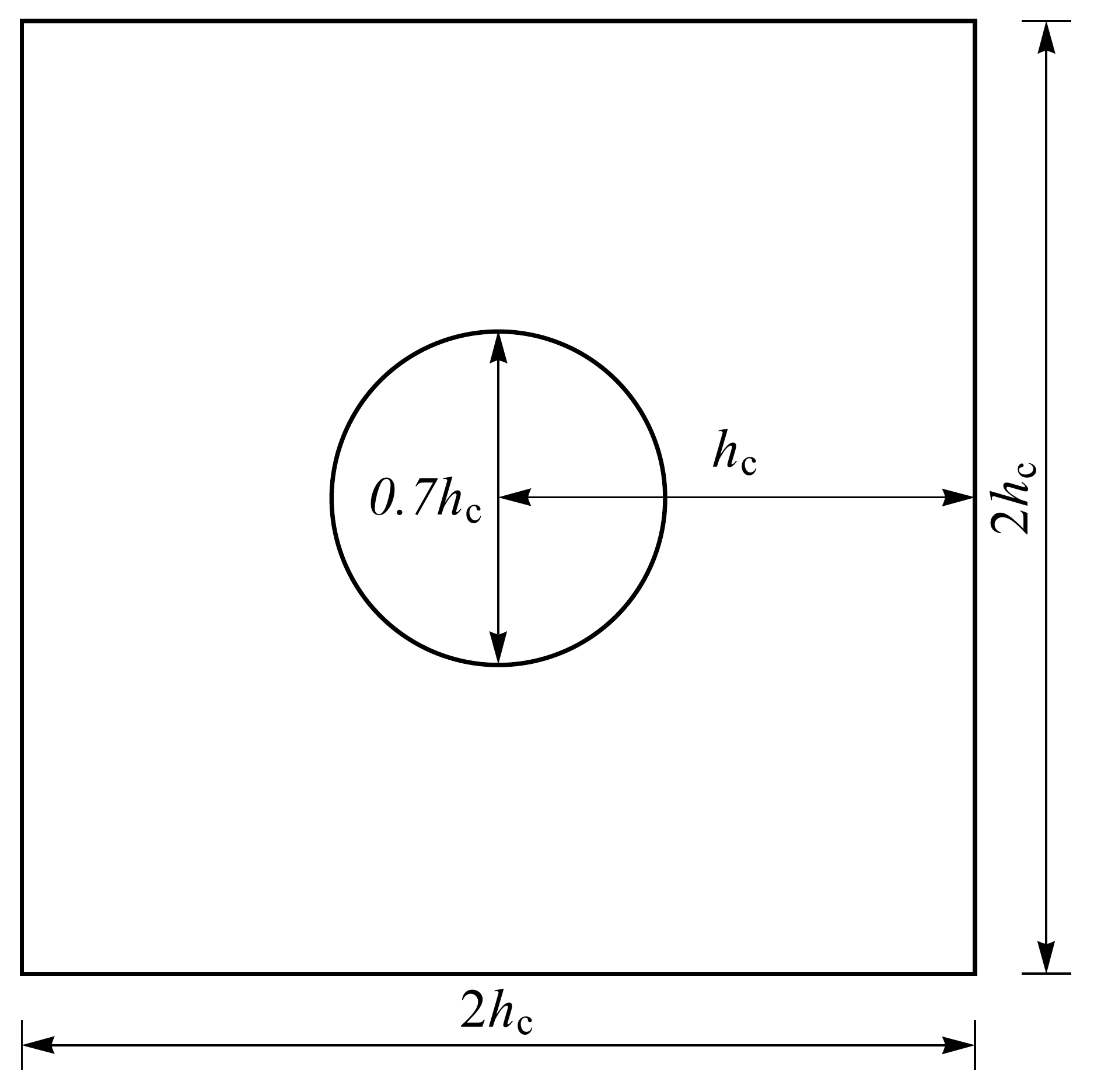}
\includegraphics[width=0.45\textwidth]{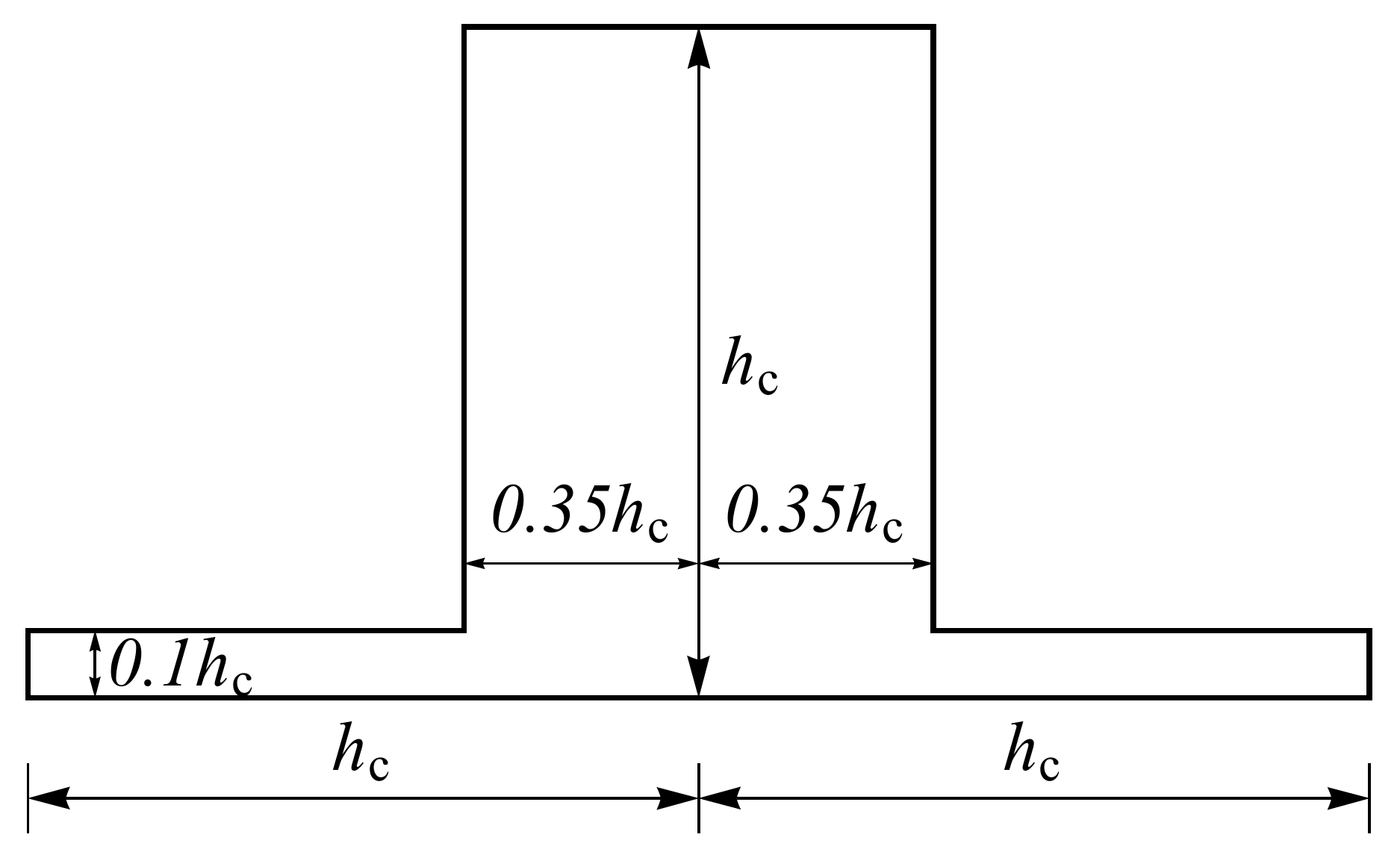}
\caption{\revise{Configuration of the 2D circular dam-break problem, left - top view, and right - side view. The illustrations are not to scale}.\label{fig:ill2d}}
\end{center}
\end{figure}

To further test the capability of the DBMPE, we employ a two-dimensional circular dam-\revise{break} problem. In this case, a cylinder water column with height $h_c$ and radius $r_c$ is set in the initial time and start to collapse with the time. For the benchmark purpose,  we choose the problem setup given by Billett and Toro\cite{billett_waf-type_1997}, where the flow is transcritical. As shown in Fig.~\ref{fig:ill2d}, by choosing $h_c$ as the reference length, the computational domain is a $2\times2$ square, and the initial radius of water column is set to be $r_c=0.35$ while the water depth in other area is set to be $0.1$. In numerical simulations, the set of 16 discrete velocities is employed together the fourth-order expansion for the equilibrium function. \revise{To adapt to the non-integer quadraure, we use a second-order upwind scheme in space for this case, while the parameters for calculation are chosen as
$\mathcal{K}=0.001$, $dx=dy=0.005$, and $dt=0.0001414$.}

\begin{figure}[hbtp!]
\begin{center}
\includegraphics[width=0.4\textwidth]{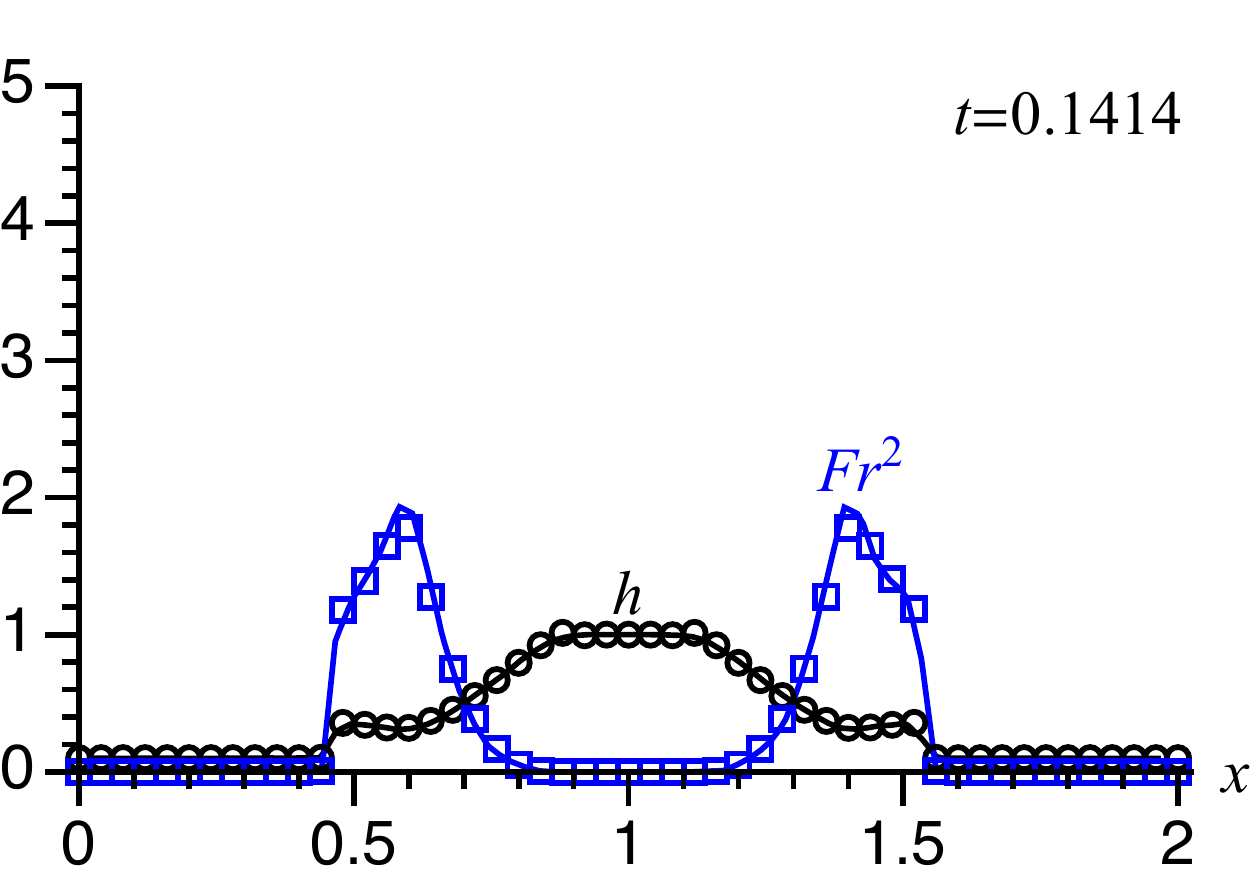}\hspace{1cm}
\includegraphics[width=0.4\textwidth]{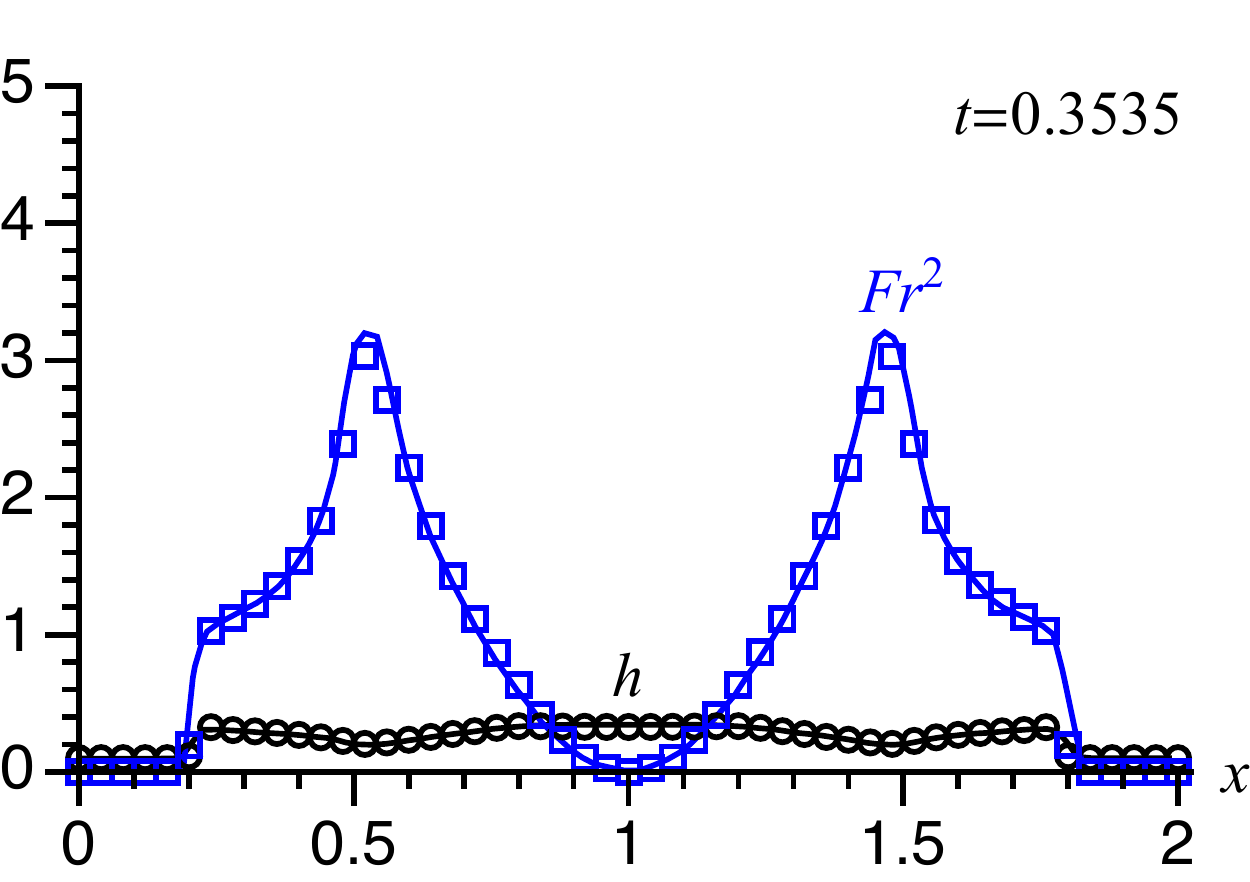}
\caption{Profiles of depth and the square of local Froude number at the $x$-strip $y=1$ for the circular dam-\revise{break} problem. Lines and symbols represent reference solutions present in \cite{billett_waf-type_1997} and the present numerical solutions, respectively.\label{fig:circular}}
\end{center}
\end{figure}

\begin{figure}[hbtp!]
\begin{center}
\includegraphics[width=0.45\textwidth]{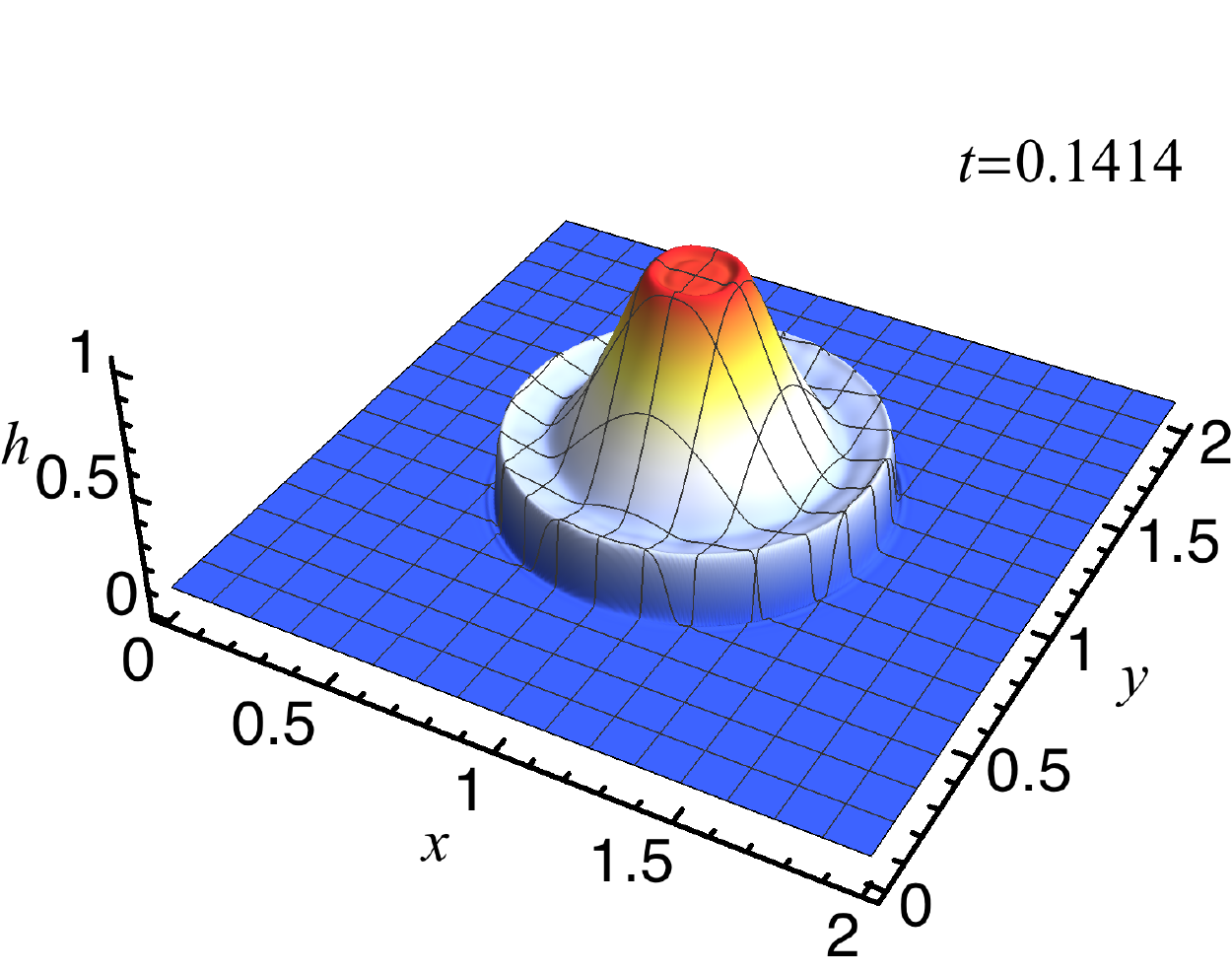}\hspace{1cm}
\includegraphics[width=0.45\textwidth]{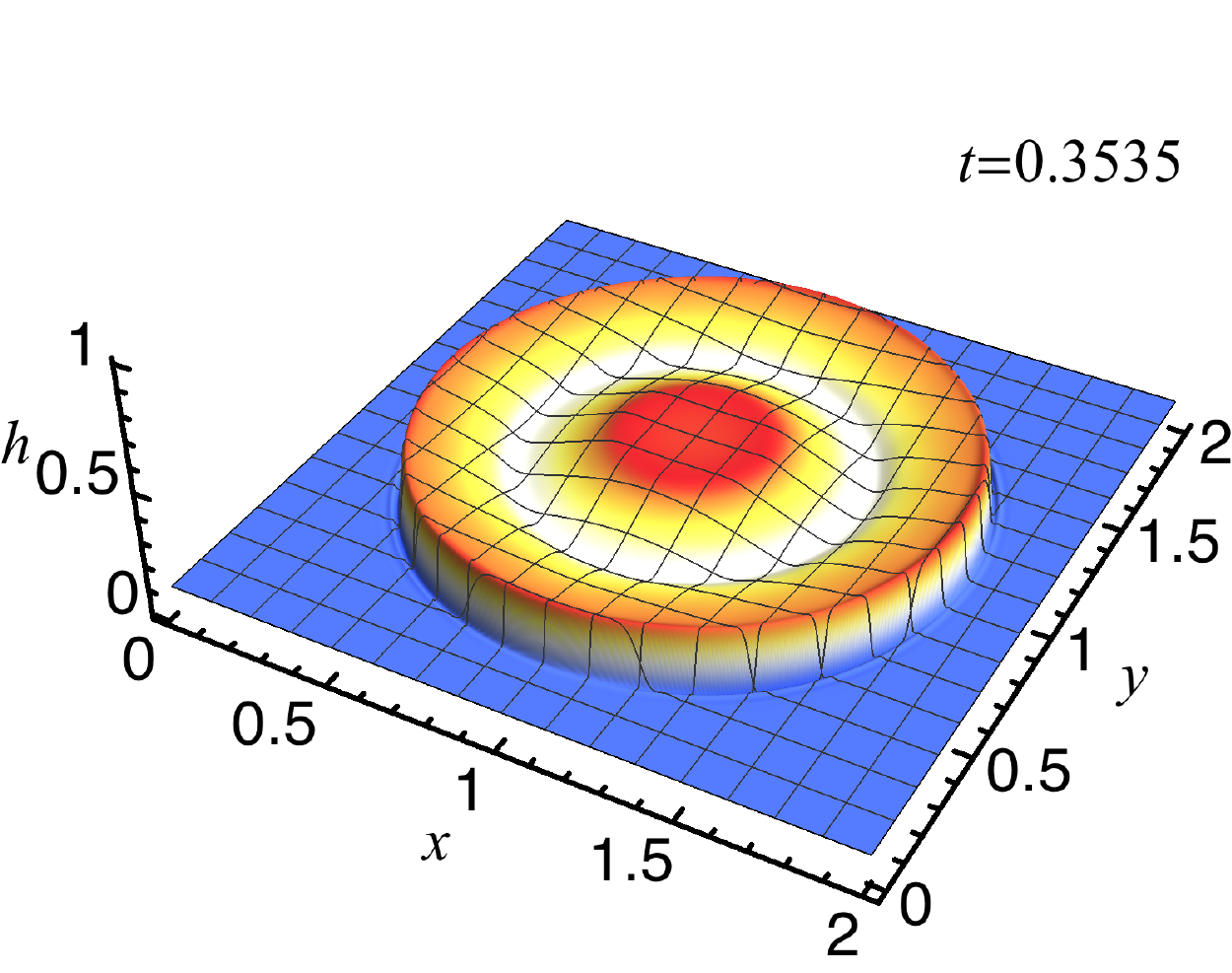}
\caption{\revise{Depth profiles for the 2D circular dam-\revise{break} problem. The coordinate lines are shown in the plots.}\label{depthprofile}}
\end{center}
\end{figure}

In Fig.~\ref{fig:circular}, the profiles of depth and local Froude number at the x-strip $y=1$ are compared with \revise{those obtained by Billett and Toro \cite{billett_waf-type_1997} using a finite volume scheme with the van Leer limiter}. As has been shown, two sets of numerical results agree well with each other. In the comparison, we notice the differences of the non-dimensional system between two works. In particular, the "$Fr$" defined in \cite{billett_waf-type_1997} is actually $Fr^2$ in this paper.  In Fig.~\ref{depthprofile}, we \revise{also}  show the depth profiles for the whole domain. Clearly the higher order expansion and  quadrature help in simulating supercritical flows.

\section{Concluding remarks}
To summarize, we have derived a type of discrete Boltzmann model
with polynomial equilibria for simulating shallow water flows. Utilizing
the Hermite expansion and the Gauss-Hermite quadrature, the DBMPE naturally
satisfies the conservation property of the collision integral, which
greatly simplifies the algorithm. We also discussed the impact of truncating distribution functions and discretizing particle velocity on the accuracy of DBMPE, and give indications on choosing the expansion order and quadrature according to the requirement of problems, e.g., $Fr$. In particular, if the quadrature consists of integer abscissae, we can obtain the very simple and efficient lattice Boltzmann scheme. Moreover, we found that there is no sacrifice in the well-balanced property for treating the source term. Further numerical validations for the one-dimensional dam-\revise{break} problem and the two-dimensional dam-\revise{break} show that the capability of simulating supercritical flows is enhanced by increasing the expansion and quadrature order.

\revise{By providing the flexibility on deriving both the single-speed DBM suitable for the stream-collision scheme, and various high-order DBMs, we expect that the applications of these models range from subcritical flows to supercritical flows. In the near future, we will further study those applications concerning supercritical flows. There are also fundamental questions remaining open, e.g., how to achieve a well-balanced scheme if taking the discretization in space and time into account, in particular for high-order models where general finite difference schemes are required.}

\section*{Acknowledgements}
Authors from the Daresbury laboratory would like to thank the Engineering and Physical Science Research Council (EPSRC) for their support under projects EP/P022243/1 and EP/R029598/1. Authors from the Sichuan University would like to thank the support of the National Natural Science Foundation of China under grant numbers: 51579166 and 5151101425.


\end{document}